\documentstyle[12pt]{article}

\begin{document}

\author{G.V. Vlasov
\thanks{
E-mail: vs@landau.ac.ru}}
\title{The superfluidity of fermions coupled to gravity}
\maketitle

\begin{abstract}
We investigate superfluidity of the relativistic fermi-gas
with
gravitational interaction.
The
exciteation spectrum is obtained within the linearized theory.
While superfluidity may take place at a definite ratio of the Fermi
momentum, rest mass and coupling constant,
the metric coefficients
play predominant role forming the gap of excitation spectrum.
\end{abstract}

\unitlength=1pt \sloppy

\section{Introduction}

The superfluidity (superconductivity, in the strict sense) of relativistic
fermi-systems is under scientific attention since Bailin and Love~\cite{BL84}
initiated investigation in this field.
Several significant papers were dedicated to superfluidity in nuclear
systems~\cite{nsf} with direct applications to a neutron star matter. Among
them the paper by Kucharek and Ring~\cite{KR91} and, especialy, Ref.~
\cite{G96} and~\cite{sf99}
should be noted as the basic works where
the general approach to relativistic superfluidity is outlined.
Besides,
Capelle and Gross~\cite{CG95} have shown how to find the excitation
spectrum.

In the present paper we shall discuss superfluidity of a fermi-system with
gravitational interaction between particles. This kind of fermi gas attracts
the interst in recent years~\cite{gravity}.

In general the investigation of superfluidity includes several steps. 1)
Specifying the Lagrangian $L$ or Hamiltonian $H$; 2) specifying the
anomalous terms there; 3) the Bogolubov transformation; 4) obtaining the
equations of motion; 5) derivation of the excitation spectrum from equations
of motion truncated up to the Hartree approximation.

\subsection{The interacting terms}

For a given Lagrangian $L$ of the many-fermion interacting system the
canonical formalism

\begin{equation}
H=L-\dot \varphi _i\pi ^i\qquad \pi ^i=\frac{\partial L}{\partial \dot
\varphi _i}  \label{canon}
\end{equation}
allows to find the Hamiltonian
\begin{equation}
H=T+U  \label{H}
\end{equation}
consisting of the kinetic energy of free fermions
\begin{equation}
T=\int {\rm d}^3{\rm r}\,\left( i\bar \psi \/\gamma ^k\partial _k\psi
\,+m\right)  \label{kin}
\end{equation}
and the interaction term
\begin{equation}
U=\int {\rm d}^3{\rm r}\,\,\bar \psi \/\Sigma \psi \,\qquad \Sigma
=\,f_i\,\,\Gamma ^iA_i  \label{U}
\end{equation}
which represents the interaction with field $A_i$\ corresponding to vertex $%
\Gamma ^i$\ and coupling constant $\,f_i$. If $\Gamma $ includes no {\it %
derivative} coupling we can extract the interacting term immediately from
the Lagrangian $L$.

For instance
\begin{equation}
\/\Gamma ^\omega =\gamma ^\mu \qquad \/\Gamma ^\sigma ={\bf 1}
\label{vertex}
\end{equation}
are the vertices of $\sigma -\omega $ model, while the vertex of QCD is
\begin{equation}
\/\Sigma =f\,\gamma ^\mu \lambda _a\/\,A_\mu ^a  \label{QCD}
\end{equation}
In order to present the Hamiltonian of great canonical ensemble, the
chemical potential~$\mu $ is introduced~as an additional term
\begin{equation}
-\mu \,\psi ^{\dagger }\/\psi  \label{chem}
\end{equation}
Indeed,
\begin{equation}
\int {\rm d}^3{\rm r\,}\mu \,\psi ^{\dagger }\/\psi \equiv \mu N  \label{N}
\end{equation}
This term stands in (\ref{U}) implicitly, or we can extract $-\mu \,$
directly from the scalar potential:
\begin{equation}
\Sigma \rightarrow \Sigma -\gamma ^0\mu  \label{sigma}
\end{equation}

The Hamiltonian of superfluid fermi-system includes besides (\ref{U}) the
anamalous term~\cite{BL84,CG95}
\begin{equation}
W=\bar \psi _c\,\beta \Delta \psi +\bar \psi _c^{\dagger }\,\Delta ^{\dagger
}\beta \psi ^{\dagger }\qquad \Delta =\,f_i\,\,\Gamma ^ia_i  \label{sf}
\end{equation}
constructed from (\ref{U}), where $a_i$ is the anamalous field (do not mix
it with $A_i$), while $\bar \psi _c=-\psi ^TC$ and $\psi _c=C\bar \psi
^T=C\beta \psi ^{*}$ are the charge-conjugated spinors and

\begin{equation}
C=i\left(
\begin{array}{cc}
\tau _2 & 0 \\
0 & \tau _2
\end{array}
\right)  \label{charge}
\end{equation}
is the matrix of charge conjugation. It is clear that
\begin{equation}
-C^T=C=\beta C\beta \qquad C^2=-1  \label{c1}
\end{equation}
\begin{equation}
C\beta =\beta C=i\left(
\begin{array}{cc}
\tau _2 & 0 \\
0 & -\tau _2
\end{array}
\right)  \label{c2}
\end{equation}
and $\bar \psi _c\psi $ is replaced by $\varphi _{\uparrow }\varphi
_{\downarrow }$ in the non-relativistic limit.

\subsection{The Bogolubov transformation}

Substituting
\begin{equation}
\hat \psi =\sum\limits_p\phi _p\left( r\right) \,\hat b_p^{\dagger }
\label{psi0}
\end{equation}
we find a common field equation:
\begin{equation}
\left\{ -i\vec \gamma \cdot \nabla +m+\Sigma \right\} \varphi _p=\beta
\varepsilon _p\varphi _p\/  \label{eqns}
\end{equation}
(no summation over $p$ in the right side) that is, briefly,
\begin{equation}
\hat h\varphi =\varepsilon \varphi \/\qquad \hat h=-i\vec \alpha \cdot
\nabla +\beta m+\beta \Sigma  \label{hamilton}
\end{equation}

As soon as the anamalous term\thinspace (\ref{sf}) appears we must use the
Bogolubov transformation
\begin{equation}
\psi =\sum\limits_pu_p\left( r\right) \,\hat b_p^{\dagger }+v_p^{*}\left(
r\right) \,\hat b_p\qquad |u_p|^2+|v_p|^2=1  \label{psi1}
\end{equation}
instead of\thinspace (\ref{psi0}). For short we shall omit index $p$; for
instance
\begin{equation}
\psi =u\,\hat b^{\dagger }+v^{*}\,\hat b\qquad \psi ^{\dagger }=u^{*}\hat
b\,+v\hat b^{\dagger }  \label{psi2}
\end{equation}
Thereby, we get two equations
\begin{equation}
\begin{array}{c}
\left\{ -i\vec \gamma \cdot \nabla +m+\Sigma \right\} u=\beta \varepsilon
u\/-\Delta ^{\dagger }\beta Cv \\
\left\{ -i\vec \gamma \cdot \nabla +m+\Sigma \right\} ^{\dagger }v=-\beta
\varepsilon v\/+C\,\beta \Delta u
\end{array}
\label{KS}
\end{equation}
for functions $u$ and $v$ instead of one for $\phi $. Therefore,
\begin{equation}
\left(
\begin{array}{cc}
\hat h+\Sigma & \beta \Delta ^{\dagger }\beta C \\
-C\,\Delta & \hat h+\Sigma
\end{array}
\right) \left(
\begin{array}{c}
u \\
v
\end{array}
\right) =E\left(
\begin{array}{c}
u \\
v
\end{array}
\right)  \label{matrix}
\end{equation}

Capelle and Gross\thinspace \cite{CG95} found two branches of excitations.
It is the analogue of the massive and acoustic modes in relativistic
bose-condensate~\cite{BY95}. The subject of our interest is the massless
mode whose excitation spectrum within the Hartree approximation looks as
\begin{equation}
E^2=\left( \varepsilon -\mu \right) ^2+|\!|\Delta |\!|^2\qquad \varepsilon
^2=\vec p^2+m^2  \label{single}
\end{equation}
or
\begin{equation}
E^2\left( \xi \right) =\xi ^2+|\!|\Delta |\!|^2\left( \xi \right)  \label{E}
\end{equation}
where the gap
\begin{equation}
|\!|\Delta |\!|^2=-C\,\Delta \beta \Delta ^{\dagger }\beta C  \label{gap}
\end{equation}
is determined immediately from~(\ref{matrix}). Particularly, for the scalar
field we get from\thinspace (\ref{gap}) the result of Capelle and
Gross\thinspace \cite{CG95}$\,\,|\!|\Delta |\!|^2=|a|^2$.

Discussion in the frames of Bogolubov-Hartree-Fock approximation, for the
scalar-vector and pseudoscalar-isovector vertices, is performed in\thinspace
\cite{KR91,G96}. The basis of exact solution in the frames of density
functional theory is not emphasized yet.

\section{Coupling with the gravitational field}

The Lagrangian of fermions coupled to the gravitational field~\cite{PS96}
reads as

\begin{equation}
\Lambda =\sqrt{-g}L\qquad g=\det g_{\mu \nu }  \label{lam}
\end{equation}
where $g_{\mu \nu }$ is the metric tensor and
\begin{equation}
L=i\bar \psi \gamma ^\mu \nabla _\mu \psi +m\bar \psi \psi  \label{L}
\end{equation}
has the form of usual free Lagrangian~containing, however, gamma-matrices $%
\gamma ^\mu $ and covariant derivative

\begin{equation}
\nabla _\mu =\partial _\mu +\Omega _\mu  \label{covar}
\end{equation}
with properties
\begin{equation}
\begin{array}{c}
\ \Omega _\mu =\frac 14\gamma _{;\mu }^\nu \,\gamma _\nu =-\frac 14\Omega
_{\mu ab}\Sigma _{ab} \\
\Omega _{\mu ab}=\partial _\mu g_{\alpha a}g^{\alpha b}-\Gamma _{\mu \beta
}^\alpha g_{\alpha a}g^{\beta b} \\
\ \Sigma _{ab}=\frac 12(\gamma _a\gamma _b-\gamma _b\gamma _a)\qquad \Sigma
_{ab}^2=1\qquad \gamma ^a\Sigma _{ab}=0
\end{array}
\label{O1}
\end{equation}
The Lagrangian includes the nonlinear terms (with respect to field $g_{\mu
\nu }$) which differ sufficiently from the usual Lagrangians in flat space
whose interacting terms are presented in a simple current-field
form\thinspace (\ref{U}).

\section{The linearized theory}

This linearized approximation~\cite{PS96}:

\begin{equation}
\begin{array}{c}
g_{\alpha a}\approx \eta _{\alpha a}+\frac 12h_{\alpha a} \\
g_{\mu \nu }=g_{\mu a}g_{\nu a}\approx \eta _{\mu \nu }+\frac 12(h_{\mu \nu
}+h_{\nu \mu }) \\
\sqrt{-g}\approx 1+\frac 12h_{aa}=1+\frac 12h \\
\Omega _{\mu ab}\approx \frac 14\left\{ \partial _\mu h_{ba}+\partial _\mu
h_{ab}-\partial _bh_{\mu a}-\partial _bh_{a\mu }+\partial _ah_{\mu
b}+\partial _ah_{b\mu }\right\} \\
\Omega _\mu \cong -\frac 18\Sigma _{ab}\partial _{[a}h_{b]\mu }\equiv \frac
14\Sigma _{ab}\partial _bh_{a\mu }
\end{array}
\label{O2}
\end{equation}
(where $\eta _{\mu \nu }$ is the Minkowsky metric) with respect to weak
field $h_{\mu \nu }$ allows to simplify Lagrangian~(\ref{L}) and split
\begin{equation}
\begin{array}{c}
L=\left[ 1+\frac 12h\right] i\bar \psi \gamma ^a\left[ \eta _a^\mu +h_a^\mu
\right] \left[ \partial _\mu +\Omega _\mu \right] \psi ++\left[ 1+\frac
12h\right] m\bar \psi \psi
\end{array}
\label{L1}
\end{equation}
into a sum
\begin{equation}
L=L_0+\tilde L  \label{L-free}
\end{equation}
of free Lagrangian
\begin{equation}
L_0=\bar \psi \left( i\bar \psi \gamma ^\mu \partial _\mu +m\right) \psi
\label{L0}
\end{equation}
and the coupling term
\begin{equation}
\tilde L=\bar \psi \left\{ i\left( \frac 12h\gamma ^\mu +\gamma ^ah_a^\mu
\right) \partial _\mu +i\gamma ^\mu \Omega _\mu +\frac 12hm\right\} \psi
\label{L-int}
\end{equation}

\section{Hamiltonian of the linearized theory}

Having
\begin{equation}
\varphi _i=\left\{ \psi ;h_{\mu \nu }\right\}  \label{variables}
\end{equation}
as dynamical variables, we find Hamiltonian~(\ref{canon}) as a sum
\begin{equation}
H=H_0+U  \label{H1}
\end{equation}
of free Hamiltonian
\begin{equation}
H_0=\bar \psi \left( i\gamma ^k\partial _k+m\right) \psi  \label{H0}
\end{equation}
and interacting term
\begin{equation}
\begin{array}{c}
U=\bar \psi \left\{ \frac 12hm+i\left( \frac 12h\,\gamma ^k+\gamma
^ah_a^k\right) \partial _k+i\gamma ^\mu \Omega _\mu \right\} \psi
\end{array}
\end{equation}
On account of derivative coupling, the term~(\ref{U}) differs from~(\ref{L}).

According to formula~(\ref{sf}) the gap matrix is
\begin{equation}
\Delta =i\left( \frac 12h\gamma ^k+\gamma ^ah_a^k\right) \partial _k+i\gamma
^\mu \Omega _\mu +\frac 12hm  \label{gap1}
\end{equation}
In general the calculation of gap~(\ref{sf}) involves tedious arithmetics
that can be omitted in several particular cases.

\section{Particular Example: a simplified metric of the rotating massive body
}

For a rotating massive body whose metric is
\begin{equation}
h_a^\mu =\frac 14h\,\delta _a^\mu \qquad h_0^k=\vec h\qquad h,\vec h\cong
{\rm const}\Rightarrow \Omega =0  \label{metric1}
\end{equation}
the coefficient~(\ref{O2})
\begin{equation}
\Omega _\mu =\frac 14\Sigma _{0j}\partial _jh_{0\mu }+\frac 14\Sigma
_{ij}\partial _jh_{i\mu }  \label{O3}
\end{equation}
vanishes. Hence, according to~(\ref{gap1}), the gap matrix is
\begin{equation}
\Delta =\left( \frac 34h\vec \gamma +\gamma ^0\vec h\right) \cdot \vec
p+\frac 12hm  \label{delta2}
\end{equation}
and
\begin{equation}
\begin{array}{c}
\Delta ^{\dagger }=\left( \frac 34h\vec \gamma -\gamma ^0\vec h\right) \cdot
\vec p+\frac 12hm \\
\,\beta \Delta ^{\dagger }\beta =-\left( \frac 34h\vec \gamma +\gamma ^0\vec
h\right) \cdot \vec p+\frac 12hm
\end{array}
\label{delta+}
\end{equation}
where
\begin{equation}
\vec p\equiv -i\nabla  \label{momentum}
\end{equation}
Thereby,
\begin{equation}
\Delta \,\beta \Delta ^{\dagger }\beta =\frac 14h^2\left[ m^2+\frac 94\vec
p^2\right] -\left( \vec h\cdot \vec p\right) ^2
\end{equation}
and, finally,
\begin{equation}
|\!|\Delta |\!|^2=-C\,\Delta \beta \Delta ^{\dagger }\beta C=\frac
14h^2\left[ m^2+\frac 94\vec p^2\right] -\left( \vec h\cdot \vec p\right) ^2
\label{gap2}
\end{equation}
Note that coefficient $\frac 94$ results from the tensor nature of
gravitational coupling: a pure scalar coupling $\bar \psi \frac 12hm\psi $
added to the free Lagrangian (\ref{L0}) leads merely to the gap
\begin{equation}
|\!|\Delta |\!|^2=\frac 14h^2m^2  \label{gap-BCS}
\end{equation}
which determines the ordinary excitation spectrum of BCS theory~\cite{BCS}
with quadratic dependence of excitation energy~(\ref{E}) on $\xi $.

Substituting the expression
\begin{equation}
\left( \xi +\mu \right) ^2-m^2=\vec p^2\qquad \mu >m  \label{xi}
\end{equation}
in~(\ref{gap2}), we get
\begin{equation}
|\!|\Delta |\!|^2\left( \xi ,\chi \right) =\frac 14h^2\left[ \frac 94\left(
\xi +\mu \right) ^2-\frac 54m^2\right] -\vec h^2\,\left[ \left( \xi +\mu
\right) ^2-m^2\right] \cos ^2\chi  \label{gap2xi}
\end{equation}
Superfluidity takes place if the right side of~(\ref{gap2xi}) is positive at
any value of $\xi $.

\section{Summary}

\subsection{Conclusion for a pure spherical metric}

Since the Fermi energy $\varepsilon _F\equiv \mu >m$, the gap~(\ref{gap2xi})
corresponding to metric of non-rotating body ($\vec h=0$) is positive at any
$\xi $ for
\begin{equation}
|\!|\Delta |\!|^2\left( \xi \right) =\frac 14h^2\left[ \frac 94\left( \xi
+\mu \right) ^2-\frac 54m^2\right]  \label{gap-sphere}
\end{equation}
implies occurrence of superfluidity. Note that Eq.~(\ref{gap-sphere}) does
not duplicate the BCS gap~(\ref{gap-BCS}) of pure scalar coupling but
immediately reduces to
\begin{equation}
|\!|\Delta |\!|^2\left( \xi \right) =\frac 14h^2\left[ m^2+\frac 92\xi
m\right] \approx \frac 14h^2m^2  \label{gap-NR}
\end{equation}
in the non-relativistic limit~$\mu \rightarrow m$. Indeed, the general
formula~(\ref{gap2xi}) also tends to~(\ref{gap-BCS}) for a non-relativistic
system. It should be noted that the ultra-relativistic, or massless,
Fermi-gas also has a non-zero gap
\begin{equation}
|\!|\Delta |\!|^2\left( \xi \right) =\frac 9{16}h^2\left( \xi +\mu \right) ^2
\label{gap-sphere-UR}
\end{equation}
due to tensor gravitational interaction.

\subsection{Conclusion for a metric with rotation}

While the gap corresponding to a pure spherical metric~(\ref{gap-sphere})
does not convey the qualitative difference from the ordinary phenomenon of
superfluidity with scalar coupling, the gravitational coupling whose metric
includes rotation (finite~$\vec h$) makes up the gap~(\ref{gap2xi})
depending both on momentum $|\vec p|$~(or~$\xi $) and angle~$\chi $ between $%
\vec p$ and~$\vec h$. The latter dependence is a specific property of the
gravitational coupling: it is not known in BCS with electromagnetic or
scalar coupling. Meanwhile, we have not considered pairing with the non-zero
orbital momentum (like~in~{\rm He-3}) wherein one may expect new
possibilities.

Eq.~(\ref{gap2xi}) implies that superfluidity may be forbidden at definite~$%
\chi $. The requirement of positive gap~$|\!|\Delta |\!|^2\left( \xi ,\chi
\right) $ at arbitrary~$\xi $ and~$\chi $ determines the condition
\begin{equation}
\frac 14h^2\left[ \frac 94\left( \xi +\mu \right) ^2-\frac 54m^2\right]
-\vec h^2\,\left[ \left( \xi +\mu \right) ^2-m^2\right] >0
\label{necessary1}
\end{equation}
or
\begin{equation}
\left( \frac 9{16}h^2-\vec h^2\,\right) \left( \xi +\mu \right) ^2+\left(
\vec h^2-\frac 5{16}h^2\right) m^2>0  \label{necessary2}
\end{equation}
necessary for the occurrence of superfluidity. It is satisfied at arbitrary~$%
\xi $ (it can be negative) if
\begin{equation}
h^2>\frac{16}9\vec h^2  \label{satisfied2}
\end{equation}
The same requirement is imposed on a massless field whose gap is
\begin{equation}
|\!|\Delta |\!|^2\left( \xi ,\chi \right) =\left( \xi +\mu \right) ^2\left(
\frac 9{16}h^2-\vec h^2\,\cos ^2\chi \right)  \label{massless}
\end{equation}
After all, we note that the gravitational coupling with metric tensor$~h_\nu
^\mu =\left\{ h_0^k\right\} $ does not admit superfluidity at all.

The excitation spectrum obtained, can applied to the calculation of
thermodynamic functions as it is used in the usual theory of
superconductivity~\cite{BCS,CG95}, i.e. internal energy density, temperature
dependence of gap etc.

\section{Appendix: Equation of motion for a non-superfluid system}

The equations of motion are

\begin{equation}
\left\{ -i\beta \gamma ^k\cdot \partial _k+\beta \left[ m+\Sigma
^{\#}\right] \right\} u_p=\epsilon _pu_p\/  \label{KSeqns}
\end{equation}
\begin{equation}
\/\left\{ -i\beta \gamma ^k\cdot \partial _k+\beta \left[ m+g_i\Gamma
^iA_i^{\#}\right] \right\} \varphi _p\left( r\right) =\epsilon _p\varphi
_p\left( r\right)  \label{KSeqns2}
\end{equation}
where we have used the notation $\hat \psi \left( r\right)
=\sum\limits_p\varphi _p\left( r\right) \hat b_p^{\dagger }\/$ with index $%
p=\left\{ p\sigma \right\} $ related to single-particle baryon densities;
also $\hat \psi \left( r\right) =\sum\limits_pu_p\left( r\right) \hat
b_p+v_p^{*}\left( r\right) \hat b_p^{\dagger }$. The self-consistent local
potentials are defined as
\begin{equation}
A_i^{\#}\left( r\right) =A_i\left( r\right) +\frac{\delta R\left[ n\right] }{%
\delta n^i\left( r\right) }\/  \label{locpot}
\end{equation}
where the interacting energy $R\left[ n\right] =E_H\left[ n\right]
+E_x\left[ n\right] +E_c\left[ n\right] $ includes the Hartree, exchange and
correlation contribution. Particularly, $A_H^{\#}\left( r\right) =\int
dtd^3r_2\/\Delta _i\left( t,r_2-r_1\right) n^i\left( r_2\right) $.

\end{document}